\documentclass[twocolumn,showpacs,aps,prl,superscriptaddress]{revtex4-1}
\usepackage{endnotes}
\usepackage{graphicx}
\usepackage{multirow}
\usepackage{hyperref}
\begin{document}

\title{Transverse Beam Spin Asymmetries at Backward Angles in Elastic Electron-Proton and Quasi-elastic Electron-Deuteron Scattering}

\author{D.~Androi\'c}
\affiliation{Department of Physics, University of Zagreb, Zagreb HR-41001 Croatia} 

\author{D.~S.~Armstrong}
\affiliation{Department of Physics, College of William and Mary, Williamsburg, VA 23187 USA} 

\author{J.~Arvieux$^\dagger$}
\affiliation{Institut de Physique Nucl\'eaire d'Orsay, Universit\'e Paris-Sud, F-91406 Orsay Cedex FRANCE}

\author{S.~L.~Bailey}
\affiliation{Department of Physics, College of William and Mary, Williamsburg, VA 23187 USA} 

\author{D.~H.~Beck}
\affiliation{Loomis Laboratory of Physics, University of Illinois, Urbana, IL 61801 USA}

\author{E.~J.~Beise}
\affiliation{Department of Physics, University of Maryland, College Park, MD 20742 USA}

\author{J.~Benesch}
\affiliation{Thomas Jefferson National Accelerator Facility, Newport News, VA 23606 USA}

\author{F.~Benmokhtar}
\affiliation{Department of Physics, University of Maryland, College Park, MD 20742 USA}
\affiliation {Department of Physics, Carnegie Mellon University, Pittsburgh, PA 15213 USA}

\author{L.~Bimbot}
\affiliation{Institut de Physique Nucl\'eaire d'Orsay, Universit\'e Paris-Sud, F-91406 Orsay Cedex FRANCE}

\author{J.~Birchall}
\affiliation{Department of Physics, University of Manitoba, Winnipeg, MB R3T 2N2 CANADA}

\author{P.~Bosted}
\affiliation{Thomas Jefferson National Accelerator Facility, Newport News, VA 23606 USA}

\author{H.~Breuer}
\affiliation{Department of Physics, University of Maryland, College Park, MD 20742 USA}

\author{C.~L.~Capuano}
\affiliation{Department of Physics, College of William and Mary, Williamsburg, VA 23187 USA} 

\author{Y.-C.~Chao}
\affiliation{Thomas Jefferson National Accelerator Facility, Newport News, VA 23606 USA}

\author{A.~Coppens}
\affiliation{Department of Physics, University of Manitoba, Winnipeg, MB R3T 2N2 CANADA} 

\author{C.~A.~Davis}
\affiliation{TRIUMF, Vancouver, BC V6T 2A3 CANADA}

\author{C.~Ellis}
\affiliation{Department of Physics, University of Maryland, College Park, MD 20742 USA}

\author{G.~Flores}
\affiliation{Department of Physics, New Mexico State University, Las Cruces, NM 88003 USA}

\author{G.~Franklin}
\affiliation {Department of Physics, Carnegie Mellon University, Pittsburgh, PA 15213 USA}

\author{C.~Furget}
\affiliation{LPSC, Universit\'e Joseph Fourier Grenoble 1, CNRS/IN2P3, Institut  Polytechnique de Grenoble, Grenoble, FRANCE}

\author{D.~Gaskell}
\affiliation{Thomas Jefferson National Accelerator Facility, Newport News, VA 23606 USA}

\author{M.~T.~W.~Gericke}
\affiliation{Department of Physics, University of Manitoba, Winnipeg, MB R3T 2N2 CANADA}

\author{J.~Grames}
\affiliation{Thomas Jefferson National Accelerator Facility, Newport News, VA 23606 USA}

\author{G.~Guillard}
\affiliation{LPSC, Universit\'e Joseph Fourier Grenoble 1, CNRS/IN2P3, Institut  Polytechnique de Grenoble, Grenoble, FRANCE}

\author{J.~Hansknecht}
\affiliation{Thomas Jefferson National Accelerator Facility, Newport News, VA 23606 USA}

\author{T.~Horn}
\affiliation{Thomas Jefferson National Accelerator Facility, Newport News, VA 23606 USA}

\author{M.~K.~Jones}
\affiliation{Thomas Jefferson National Accelerator Facility, Newport News, VA 23606 USA}

\author{P.~M.~King}
\affiliation{Department of Physics and Astronomy, Ohio University, Athens, OH 45701 USA}

\author{W.~Korsch}
\affiliation{Department of Physics and Astronomy, University of Kentucky, Lexington, KY 40506 USA}

\author{S.~Kox}
\affiliation{LPSC, Universit\'e Joseph Fourier Grenoble 1, CNRS/IN2P3, Institut  Polytechnique de Grenoble, Grenoble, FRANCE}

\author{L.~Lee}
\affiliation{Department of Physics, University of Manitoba, Winnipeg, MB R3T 2N2 CANADA}

\author{J.~Liu}
\affiliation{Kellogg Radiation Laboratory, California Institute of Technology,  Pasadena, CA 91125 USA}

\author{A.~Lung}
\affiliation{Thomas Jefferson National Accelerator Facility, Newport News, VA 23606 USA}

\author{J.~Mammei}
\affiliation{Department of Physics, Virginia Tech, Blacksburg, VA 24061 USA}

\author{J.~W.~Martin}
\affiliation{Department of Physics, University of Winnipeg, Winnipeg, MB R3B 2E9 CANADA}

\author{R.~D.~McKeown}
\affiliation{Kellogg Radiation Laboratory, California Institute of Technology,  Pasadena, CA 91125 USA}

\author{A.~Micherdzinska}
\affiliation{Department of Physics, The George Washington University, Washington, DC 20052 USA}

\author{M.~Mihovilovic}
\affiliation{Jo\^zef Stefan Institute, 1000 Ljubljana, SLOVENIA}

\author{H.~Mkrtchyan}
\affiliation{Yerevan Physics Institute, Yerevan 375036 ARMENIA}

\author{M.~Muether}
\affiliation{Loomis Laboratory of Physics, University of Illinois, Urbana, IL 61801 USA}

\author{S.~A.~Page}
\affiliation{Department of Physics, University of Manitoba, Winnipeg, MB R3T 2N2 CANADA}

\author{V.~Papavassiliou}
\affiliation{Department of Physics, New Mexico State University, Las Cruces, NM 88003 USA}

\author{S.~F.~Pate}
\affiliation{Department of Physics, New Mexico State University, Las Cruces, NM 88003 USA}

\author{S.~K.~Phillips}
\affiliation{Department of Physics, College of William and Mary, Williamsburg, VA 23187 USA} 

\author{P. Pillot}
\affiliation{LPSC, Universit\'e Joseph Fourier Grenoble 1, CNRS/IN2P3, Institut  Polytechnique de Grenoble, Grenoble, FRANCE}

\author{M.~L.~Pitt}
\affiliation{Department of Physics, Virginia Tech, Blacksburg, VA 24061 USA}

\author{M.~Poelker}
\affiliation{Thomas Jefferson National Accelerator Facility, Newport News, VA 23606 USA}

\author{B.~Quinn}
\affiliation {Department of Physics, Carnegie Mellon University, Pittsburgh, PA 15213 USA}

\author{W.~D.~Ramsay}
\affiliation{Department of Physics, University of Manitoba, Winnipeg, MB R3T 2N2 CANADA}

\author{J.-S.~Real}
\affiliation{LPSC, Universit\'e Joseph Fourier Grenoble 1, CNRS/IN2P3, Institut  Polytechnique de Grenoble, Grenoble, FRANCE}

\author{J.~Roche}
\affiliation{Department of Physics and Astronomy, Ohio University, Athens, OH 45701 USA}

\author{P.~Roos}
\affiliation{Department of Physics, University of Maryland, College Park, MD 20742 USA}

\author{J.~Schaub}
\affiliation{Department of Physics, New Mexico State University, Las Cruces, NM 88003 USA}

\author{T.~Seva}
\affiliation{Department of Physics, University of Zagreb, Zagreb HR-41001 Croatia} 

\author{N.~Simicevic}
\affiliation{Department of Physics, Louisiana Tech University,  Ruston, LA 71272 USA}

\author{G.~R.~Smith}
\affiliation{Thomas Jefferson National Accelerator Facility, Newport News, VA 23606 USA}

\author{D.~T.~Spayde}
\affiliation{Department of Physics, Hendrix College, Conway, AR 72032 USA}

\author{M.~Stutzman}
\affiliation{Thomas Jefferson National Accelerator Facility, Newport News, VA 23606 USA}

\author{R.~Suleiman}
\affiliation{Department of Physics, Virginia Tech, Blacksburg, VA 24061 USA}
\affiliation{Thomas Jefferson National Accelerator Facility, Newport News, VA 23606 USA}

\author{V.~Tadevosyan}
\affiliation{Yerevan Physics Institute, Yerevan 375036 ARMENIA}

\author{W.~T.~H.~van~Oers}
\affiliation{Department of Physics, University of Manitoba, Winnipeg, MB R3T 2N2 CANADA}

\author{M.~Versteegen}
\affiliation{LPSC, Universit\'e Joseph Fourier Grenoble 1, CNRS/IN2P3, Institut  Polytechnique de Grenoble, Grenoble, FRANCE}

\author{E.~Voutier}
\affiliation{LPSC, Universit\'e Joseph Fourier Grenoble 1, CNRS/IN2P3, Institut  Polytechnique de Grenoble, Grenoble, FRANCE}

\author{W.~Vulcan}
\affiliation{Thomas Jefferson National Accelerator Facility, Newport News, VA 23606 USA}

\author{S.~P.~Wells}
\affiliation{Department of Physics, Louisiana Tech University,  Ruston, LA 71272 USA}

\author{S.~E.~Williamson}
\affiliation{Loomis Laboratory of Physics, University of Illinois, Urbana, IL 61801 USA}

\author{S.~A.~Wood}
\affiliation{Thomas Jefferson National Accelerator Facility, Newport News, VA 23606 USA}

\collaboration{G0 Collaboration}

\author{B.~Pasquini}
\affiliation{Dipartimento di Fisica Nucleare e Teorica, Universit\`a degli
Studi di Pavia and INFN, Sezione di Pavia, Pavia, Italy} 

\author{M.~Vanderhaeghen}
\affiliation{Institut f\"ur Kernphysik, Johannes Gutenberg Universit\"at, D-55099
Mainz, Germany} 

\date{\today}

\begin{abstract}
We have measured the beam-normal single-spin asymmetries in elastic scattering of transversely polarized electrons from the proton, and performed the first measurement in quasi-elastic scattering on the deuteron, at backward angles (lab scattering angle of 108$^{\circ}$) for Q$^2$ = 0.22~GeV$^2/c^2$ and 0.63~GeV$^2/c^2$ at beam energies of 362~MeV and 687~MeV, respectively.  The asymmetry arises due to the imaginary part of the interference of the two-photon exchange amplitude with that of single-photon exchange.  Results for the proton are consistent with a model calculation which includes inelastic intermediate hadronic ($\pi$N) states.  An estimate of the beam-normal single-spin asymmetry for the scattering from the neutron is made using a quasi-static deuterium approximation, and is also in agreement with theory.
\end{abstract}

\maketitle

Two-photon exchange (TPE) is a higher-order radiative effect that may explain the discrepancy between different methods of measuring the ratio of the electric and magnetic form factors of the proton ($G_E$ and $G_M$) \cite{GV2003}.  Model calculations have shown that including the real part of TPE effects brings the unpolarized cross-section measurements into closer agreement with polarization transfer measurements \cite{arrmeltjon2007}.  The contribution of higher-order processes, such as TPE effects, must be understood as the precision of electron scattering experiments continues to improve \cite{Tjon09}.  The single-spin asymmetry in electron-nucleon scattering (the left-right analyzing power measured in the p($\vec{e}$,e)p reaction, with the sign chosen as in the Madison convention \cite{madison}) as measured in this work is a parity-conserving asymmetry which gives access to the imaginary part of TPE, providing a valuable test of the theoretical framework for such higher-order processes.  It can be measured with either the target or the beam polarized perpendicular (transverse) to the scattering plane \cite{transversetheory}.  

The asymmetry arises due to the imaginary part of the interference of the two-photon exchange amplitude with that of single-photon exchange.  The beam-normal single-spin (BNSSA) asymmetry, $B_n$, can be written \begin{equation}B_n=\frac{\sigma_{\uparrow}-\sigma_{\downarrow}}{\sigma_{\uparrow}+\sigma_{\downarrow}}=\frac{2\,{\rm Im}(M^*_\gamma\,\cdot\,\left|M_{\gamma\gamma}\right|)}{|M_\gamma|^2}\label{transasym}\end{equation} where Im denotes the imaginary part, $\sigma_{\uparrow}$ and $\sigma_{\downarrow}$ are the cross-sections for the beam polarized parallel or antiparallel to the normal to the scattering plane, $\hat{n}=\frac{\vec{k}\times\vec{k^{\prime}}}{\left|\vec{k}\times\vec{k^{\prime}}\right|}$ where $\vec{k}$ and $\vec{k^{\prime}}$ are the momenta of the incoming and outgoing electron, and $M_{\gamma}$ and $\left|M_{\gamma \gamma}\right|$ are the amplitudes for single- and two- photon exchange.  The BNSSA is linear in $\alpha_{EM}$, the electromagnetic (EM) coupling constant, because the single-spin asymmetry is zero in the Born approximation.  In addition, the BNSSA has an order $m_e/E$ suppression relative to a target-normal single-spin asymmetry, where $m_e$ is the mass of the electron and E is the energy, because the polarized electron is ultra-relativistic.  The BNSSA are expected to be on the order of $10^{-6} - 10^{-5}$ \cite{transversetheory}.  

The techniques developed for measuring the small parity-violating (PV) asymmetries ($\sim$ $10^{-5}$), using an electron beam spin polarized in the same direction as the momentum (longitudinal), are also useful for measuring the BNSSA.  Precision electroweak electron scattering experiments, such as HAPPEx \cite{happextrans}, PVA4 \cite{Maas1}, E158 \cite{v2007} and G0\cite{arm2007} have measured the BNSSA at various kinematic settings at forward angles, and backward angle measurements have been made in PVA4 \cite{pva4trans, Maasprivate} (preliminary), SAMPLE \cite{sample3}, and this work.  These experiments deliberately polarize the beam in the transverse direction in order to estimate the systematic uncertainty on the PV asymmetries that could be caused by a residual transverse component of the beam and also directly provide access to the imaginary part of the TPE amplitude.

Calculations of the BNSSA are sensitive to the treatment of the intermediate hadronic state in the two-photon exchange amplitude, and different models have been used in the different kinematic regimes.  The cross section, $\sigma$, can be parameterized using 6 complex invariant amplitudes ($\tilde{G}_E(\nu,Q^2)$, $\tilde{G}_M(\nu,Q^2)$ and $\tilde{F}_i(\nu,Q^2)$ where $i=3-6$) which are functions of the four-momentum transfer $Q^2$ and the Lorentz-invariant, $\nu=(s-u)/4$, where $s$ and $u$ are the standard Mandelstam variables \cite{gor2004}.  In the Born approximation, two of the complex form factors reduce to the familiar electric and magnetic form factors $\tilde{G}_E(\nu,Q^2)\rightarrow G_E(Q^2)$, $\tilde{G}_M(\nu,Q^2)\rightarrow G_M(Q^2)$ while the remaining form factors, which originate from processes involving the exchange of at least two photons, vanish ($\tilde{F}_i(\nu,Q^2)\rightarrow 0$).  The most relevant models for the backward-angle measurements, including this work, are those which model the nucleon intermediate states, $X$, including elastic ($X=N$) and inelastic ($X=\pi N$) states.  

The results presented here are from backward angle ($\theta_{lab}$=108$^{\circ}$) measurements of the BNSSA in elastic electron scattering on hydrogen and quasi-elastic electron scattering on deuterium, taken as part of the G0 experiment  \cite{G0NIM, backprl}.  Two incident beam energies, 362~MeV and 687~MeV, were used for each target, corresponding to $Q^2 \sim$ 0.22~GeV$^2/c^2$ and 0.63~GeV$^2/c^2$, respectively.  Beam currents ranged from 20 $\mu$A to 60 $\mu$A, for a total of about 50 hours of beam.  The experimental apparatus consisted of a 20 cm aluminum target cell which was used to hold either liquid hydrogen or deuterium, and a toroidal-field magnetic spectrometer which was used to separate the (quasi-)elastically and inelastically scattered electrons.  The apparatus had eight-fold azimuthal symmetry around the beamline, with three sets of main detectors in each octant.  Focal plane detectors (FPDs), consisting of 14 scintillator arcs in each octant, and cryostat exit detectors (CEDs), an array of 9 scintillator paddles, were used for kinematic separation by looking at the coincidence of individual CED and FPD pairs.  An aerogel \v{C}erenkov detector in each octant was used to distinguish electrons and pions.  A set of synthetic quartz \v{C}erenkov luminosity (LUMI) monitors placed symmetrically around the beamline at low angles (high incident rate) was used as a beam diagnostic, as will be discussed below.  

The polarized electrons were produced from circularly polarized light incident on a strained GaAs photocathode \cite{beamref}.  Rapid helicity reversal, at 30 Hz using a Pockels cell, ensured that the conditions for which an asymmetry is measured do not change.  A slow helicity reversal ($\sim$~several days), using an insertable half-wave plate, was also employed to reduce helicity-correlated effects.  A Wien filter was used to produce a transversely polarized electron beam.  The electron polarization vector was rotated by 90$^{\circ}$ to beam left from what it would be for longitudinal polarization (as determined by Moller polarimeter measurements at various Wien angle settings).  The magnitude of the polarization of the beam was 85.8$\%$ with uncertainties of $\pm$2.0$\%$ and $\pm$1.4$\%$ for the 362~MeV and 687~MeV energies, respectively.

\begin{figure}
\includegraphics[width=250pt]{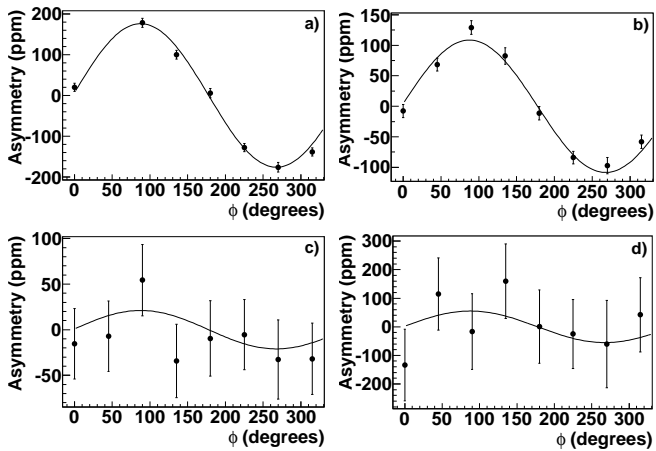}%
\caption{Measured asymmetries as a function of $\phi$ for 362~MeV from hydrogen (a) and deuterium (b), and 687~MeV from hydrogen (c) and deuterium (d). Error bars include statistical and systematic uncertainties.  Data are corrected for the magnitude of the polarization.\label{fig:asyms}}
\end{figure}

The direction of the beam polarization was flipped in a quartet pattern ($+--+$ or $-++-$) with plus (minus) corresponding to beam left (right) looking downstream.  The measured asymmetry in each octant was formed from the difference in normalized yields, $Y$, for the plus and minus states in each quartet over the sum \begin{equation}A^{\perp}_{meas} =\frac{Y_{+}-Y_{-}}{Y_{+}+Y_{-}}=B_n \vec{p}_e\cdot\hat{n}=-B_n |\vec{p}_e| sin (\phi+\phi_o)\label{transasym_2}\end{equation} where $\vec{p}_e$ is the beam polarization.  The normal to the scattering plane, $\hat{n}$, is transverse to the beam and to the scattered electron momentum, thus the measured asymmetry varies as a function of octant.  To extract the value of the transverse asymmetry, we fit the data as a sinusoidal function of azimuthal scattering angle, $\phi$, and obtained the amplitude, $B_n$, corrected for the magnitude of the beam polarization, $\left|\vec{p}_e\right|$.  

\begin{table}\caption{Summary of the fit parameters for each dataset.}
\begin{center}	
\renewcommand{\arraystretch}{1.5}{
\begin{tabular*}{0.50\textwidth}{@{\extracolsep{\fill}} c c c c c}
\hline
 & Dataset  & $B_n^p$ or $B_n^d$ (ppm)      & $\frac{\chi^2}{ndf}$  & \\ \hline
 &   H362    & -176.5 $\pm$ 9.4  & 1.6  & \\
 &   D362    & -108.6 $\pm$ 7.2  & 1.4  & \\
 &   H687    & -21.0  $\pm$ 24   & 0.4  & \\
 &   D687    & -55.7  $\pm$ 78   & 0.4  & \\ \hline
\end{tabular*}}\end{center}\label{tbl:transasyms}\end{table}

As the beam passes through the magnetic elements of the accelerator, the electron spin precesses in the horizontal plane.  The out-of-plane component of the polarization induced by these elements is small and thus is not expected to contribute significantly to the phase of the asymmetry.  The phases as determined from the main detectors are consistent from dataset to dataset, though with large uncertainties, especially in the high energy data.  To confirm the stability of the out-of-plane phase from dataset to dataset, we used the high precision LUMI data, which are dominated by M\o ller (electron-electron) scattering.  We discovered that although the LUMI phases are consistent from dataset to dataset, they were not consistent with the phases determined from the main detector data, indicating that there was a geometrical offset between the two sets of detectors \cite{juliette}.  In the final fits to the main detector data (see Figure \ref{fig:asyms}) only the amplitude was allowed to vary and the phases were fixed to the weighted average of the phases of the main detectors at 362~MeV, or $-2.3^\circ \pm 1.6^\circ$, where the data are more precise.  The results are summarized in Table \ref{tbl:transasyms}.  The uncertainties in the 362~MeV data are small compared to the size of the asymmetries and the quality of the data is apparent in the plots.  The hydrogen data at 362~MeV were missing an electronics channel in octant 2, so that octant has been omitted from the final results.  At 687~MeV both the expected values of the asymmetries and the rates were smaller, and there were very few data taken with a deuterium target, resulting in much larger relative uncertainties.  Both the hydrogen and deuterium asymmetry at 687~MeV are consistent with zero.

\begin{table}\caption{Estimates of the various contributions (statistics and corrections for electronics effects, helicity-corellated beam properties, backgrounds and polarization) to the uncertainties in each dataset.  If a source of systematic uncertainty has a contribution that is global, it is listed in parentheses.}\begin{center}	
\renewcommand{\arraystretch}{1.5}{
\begin{tabular*}{0.50\textwidth}{@{\extracolsep{\fill}}c c c c c}  \hline
 \multirow{2}{*}{Contribution}   & \multicolumn{4}{c}{Uncertainty (ppm)} \\
                               & D687  & H687 & D362 & H362       \\ \hline
 Statistics                    & 57.8  & 17.0 & 5.6  & 5.4  \\
 Electronics                   & 42.0  & 4.6  & 3.5  &  1.2  \\
 Beam Properties               &  6.0  & 2.4  & 1.6  &  2.0  \\
 Backgrounds                   & 30.7  & 15.2 (2.5) & 0.7  &  3.7 (5.1) \\
 Polarization                  &  0.6 (0.6)  & 0.2 (0.2) & 2.1 (1.3) &  3.4 (2.0)  \\
  \hline
\end{tabular*}}\end{center}\label{tbl:transuncert}\end{table}

The transverse data have been fully corrected for electronics effects (e.g. deadtime), background asymmetries and helicity-correlated beam parameters.  The corrections to the data are performed on an octant-by-octant basis in the same way as in the longitudinal data \cite{backprl}.  The uncertainties associated with each correction (see Table \ref{tbl:transuncert}) are calculated as the quadrature difference in the uncertainties on the amplitudes of the fits before and after a given correction, except in the case of the linear regression correction in the 687~MeV data, which resulted in a smaller fit uncertainty after the correction.  In this case, the uncertainty is approximated by scaling the uncertainties for the 362~MeV data by the square root of the ratios of the number of quartets in each dataset.  As there is yet no prescription to calculate the radiative effects for the beam-normal single-spin asymmetry, we have not made any corrections for standard radiative (real photon) effects \cite{MoTsai}.

\begin{figure}
\includegraphics[width=250pt]{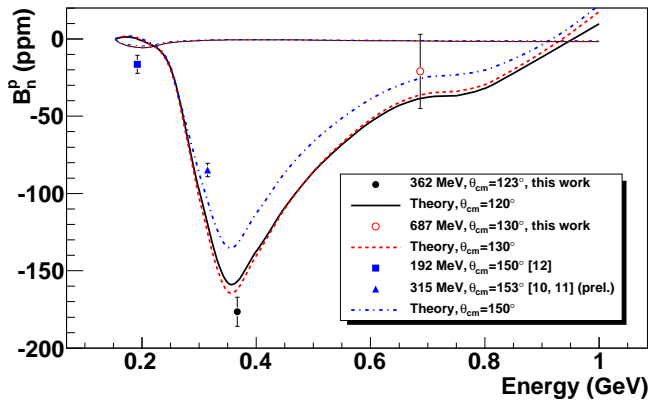}%
\caption{(color online) World data on the BNSSA at backward angles for different center-of-mass angles as a function of beam energy.  Theory curves include both the elastic and $\pi$N intermediate state contributions to the asymmetry \cite{transversetheory}.  For comparison, the purely elastic contributions are also shown (overlapping curves at approximately zero for the entire range).  
\label{fig:world_data}}
\end{figure}

\begin{table}\caption{Estimate of the proton and neutron cross sections and asymmetries for each energy, assuming a 5\% uncertainty on the cross-sections.  The theory prediction \cite{transversetheory} is given in the last column, where for the neutron it is a calculation at the exact kinematics; for the proton it is an estimate based on the curves shown in Figure \ref{fig:world_data}.}\begin{center}	
\renewcommand{\arraystretch}{1.5}{
\begin{tabular*}{0.50\textwidth}{@{\extracolsep{\fill}}c r c c c}  \hline
  Energy              & \multicolumn{2}{c}{Cross Section} & $B_n^{n,p}$       & ${B_{n, theory}^{n,p}}$   \\
   (MeV)              & \multicolumn{2}{c}{($\mu$b/sr)}   & (ppm)             & (ppm)                    \\ \hline
 \multirow{2}{*}{362} &  n  &  8                 &   86.6 $\pm$  41  &   72  \\
                      &  p  &  23               & -176.5 $\pm$ 9.4  & -158    \\
 \multirow{2}{*}{687 }&  n  &  1.1            & -138   $\pm$ 268  &  20  \\
                      &  p  &  2.6            &  -21.0 $\pm$  24  &  -35    \\
  \hline
\end{tabular*}}\end{center}\label{tbl:transxsec}\end{table} 
Our two measurements of the BNSSA for scattering from the proton are shown on a plot with the preliminary PVA4 \cite{pva4trans, Maasprivate} and the SAMPLE \cite{sample3} backward angle measurements (see Figure \ref{fig:world_data}).  The data are shown in comparison to the theoretical prediction \cite{transversetheory}.  For the first time we have extracted values for the BNSSA of the neutron.  In the static approximation, the asymmetry for deuterium is simply the cross-section-weighted average asymmetry for the proton and the neutron \begin{equation}B_n^d = \frac{\sigma_p B_n^p + \sigma_n B_n^n}{\sigma_p+\sigma_n}\label{eq:stattrans}\end{equation} where $\sigma_{p, n}$ is the proton ($p$) or neutron ($n$) cross-section, and $B_n^{n, p, d}$ is the measured BNSSA for a neutron ($n$), proton ($p$) or deuteron ($d$) target.  Estimates of the proton and neutron cross-sections and the extracted BNSSA for the neutron are given in Table \ref{tbl:transxsec} and compared to the theory \cite{G0web}.  The cross-sections were calculated using estimates of the nucleon EM form factors with a relative uncertainty of 5\%.   

The estimate for the neutron asymmetry for each energy is made by solving for $B_n^n$ in Eq. \ref{eq:stattrans}.  The estimate of the neutron BNSSA at 687~MeV has very large uncertainties which prevent us from drawing any conclusions.  At 362~MeV, the resulting neutron asymmetry is smaller in magnitude than the proton asymmetry and opposite in sign (positive).  In the resonance region the elastic contribution is calculated using the electromagnetic form factors at the vertices, while the contribution from $\pi$N intermediate states depends on both resonant and nonresonant invariant amplitudes for $\pi$N intermediate states, which are taken from phenomenological analysis fitted to available experimental data \cite{transversetheory, MAID2003}.  The asymmetry at the measured values of $Q^2$ is dominated by the term proportional to $G_M$ which changes sign between proton and neutron.  Furthermore the larger magnitude of the neutron asymmetry for smaller energies follows from the dominance of the quasi-real Compton contribution.  It corresponds to the two exchanged photons being quasi-real and the invariant mass of the hadronic intermediate state approaching the value of the \textit{e-N} center of mass energy.  In Fig. \ref{fig:world_data}, the behavior of the proton asymmetry is driven by the increasing contribution of the quasi-real Compton scattering up to energy $E_e\approx0.360$~GeV. At higher energy the resonant structure of the pion electroproduction amplitudes comes into play with a contribution of opposite sign, which leads to a smaller asymmetry in absolute value.  In order to make a better estimate of the neutron asymmetry it will be necessary to use a more sophisticated deuterium model, similar to the calculation of Schiavilla \cite{rocco1, rocco2} for the estimate of the longitudinal asymmetries.

Measurements of the BNSSA in the resonance region are valuable tests of the theoretical framework which calculates the radiative corrections for precision electron scattering experiments.  This work doubles the world dataset for the BNSSA in elastic electron-proton scattering at backward angles.  More importantly, the addition of these data allows us to span the range of energies up to 1~GeV, including the value at 362~MeV which is at the estimated peak of the theoretical prediction.  In addition, asymmetries from quasi-elastic deuteron scattering have been used to provide the first estimate of the BNSSA for the neutron, which is in agreement with the predicted value at 362~MeV.  The agreement between the theoretical predictions and the measured values clearly shows that it is necessary to take into account the $\pi$N intermediate state contributions in the calculation of the hadronic intermediate state when estimating the effects of the TPE contributions.  

\begin{acknowledgments}
We gratefully acknowledge the strong technical contributions to this experiment from many groups: Caltech, Illinois, LPSC-Grenoble, IPN-Orsay, TRIUMF and particularly the Accelerator and Hall C groups at Jefferson Lab. CNRS (France), DOE (U.S.), NSERC (Canada) and NSF (U.S.) supported this work in part.   
\end{acknowledgments}

%

\end{document}